\documentclass[aps,prl,reprint,groupedaddress]{revtex4-1}

\usepackage{graphicx}
\usepackage{subfigure}
\usepackage{comment}
\usepackage{verbatim}
\usepackage{url}

\usepackage{natbib}
\begin{document}


\title{Demonstration of Cathode Emittance Dominated High Bunch Charge Beams in a DC gun-based Photoinjector}


\author{Colwyn Gulliford}\homepage{co-first author}\email{cg248@cornell.edu}
\author{Adam Bartnik}\homepage{co-first author}\email{acb20@cornell.edu}
\author{Ivan Bazarov}
\author{Bruce Dunham}
\author{Luca Cultrera}
\affiliation{CLASSE, Cornell University}


\date{\today}

\begin{abstract}

We present the results of transverse emittance and longitudinal current profile measurements of high bunch charge ($\geq$100 pC) beams produced in the DC gun-based Cornell Energy Recovery Linac Photoinjector.  In particular, we show that the cathode thermal and core beam emittances dominate the final 95\% and core emittance measured at 9-9.5 MeV.  Additionally, we demonstrate excellent agreement between optimized 3D space charge simulations and measurement, and show that the quality of the transverse laser distribution limits the optimal simulated and measured emittances.  These results, previously thought achievable only with RF guns, demonstrate that DC gun based photoinjectors are capable of delivering beams with sufficient single bunch charge and beam quality suitable for many current and next generation accelerator projects such as Energy Recovery Linacs (ERLs) and Free Electron Lasers (FELs).  
\end{abstract}

\pacs{PACS numbers?}


\maketitle

Linear electron accelerators boast a wide range of current and planned applications in the physical sciences.  Examples include: x-ray sources \cite{ref:TJNAF,ref:pddr,ref:lcls}, electron-ion coolers \cite{ref:eRHIC}, Ultra-fast Electron Diffraction (UED) experiments \cite{ref:uedslac1,ref:uedutor1,ref:uedutor2,ref:uedbnl1}, and fixed-target nuclear physics experiments \cite{ref:mixingangle1}. A key feature of many of these applications is the potential to produce beams where the initial beam quality, set by the source, dominates the final beam quality at the usage point.  
 This has lead to the design of a next generation of machines, such as high energy Energy Recovery Linacs (ERLs) \cite{ref:pddr}, and Free Electron Lasers (FELs) \cite{ref:lcls}  which could provide diffraction limited hard x-rays with orders of magnitude brighter beams than modern storage rings.  
The successful design and implementation of such machines has the potential to impact an impressively broad range of research in physics, chemistry, biology, and engineering.  

For next generation high energy x-ray sources like the proposed Linac Coherent Light Source-II (LCLS-II) \cite{ref:slacspecs1}, the creation (at MHz repetition rates) and effective transport of multi-MeV beams with high bunch charges ($\geq$100 pC), picosecond bunch lengths, and sub-micron normalized transverse emittances represents a beam dynamics regime previously thought attainable only with RF gun based photoinjectors \cite{ref:LBNL}.  In this letter, we challenge this assumption, and show that the DC gun-based Cornell ERL injector can produce cathode emittance dominated beams which meet the bunch charge, emittance, and peak current specifications of a next generation light source. 
In doing so, we also demonstrate excellent agreement between measurement and simulation of the injector, and show that ultimate optimization of the emittance in high-brightness photoinjectors may require advanced transverse laser shaping along with the use of low intrinsic emittance photocathodes.

Before discussing our experimental results, we review the definitions of the key figures of merit for beam quality in high-brightness accelerators relevant for this work: emittance and brightness.
For the beam densities encountered in this work ($10^{17}$-$10^{18}$ $\mathrm{e}/\mathrm{m}^3)$, classical relativistic Hamiltonian mechanics, with a self-interaction described by a space charge potential, sufficiently approximates the single bunch beam dynamics \cite{ref:rsecomp,ref:disorderedheating}.  In this model, the Hamiltonian for each bunch separates into a sum over $N=q/e$ Hamiltonians of the same form, reducing the 6N-D ensemble phase space volume conserved in Liouville's theorem to the 6D phase space of a single bunch.
In the absence of coupling between each of the 2D canonical phase spaces ($x_i$, $P_{x_i}$), the conserved 6D phase space volume separates into three conserved 2D volumes.  From a physics stand point, these three conserved quantities represent the most fundamental definition of 2D emittance.  However, as a figure of merit, this definition fails to capture the effect of distortions of the phase space due to non-linear fields.  This typically motivates the definition of the (normalized) rms emittance: 
$\epsilon_{n,x_i} = \frac{1}{mc}\sqrt{\langle x_i^2\rangle\langle p_{x_i}^2\rangle-\langle x_ip_{x_i}\rangle^2}$.
Note the use of the mechanical momenta $p_{x_i}$.  Under the above assumptions, conservation of this emittance follows directly from Liouville's theorem, provided the forces on the bunch are linear.

In practice, bunches in high-brightness photoinjectors experience both non-linear and longitudinally correlated fields arising from space charge and time-dependent RF fields.  Left unchecked, these fields lead to emittance growth along the beamline.  The mitigation of these effects, known as emittance compensation \cite{ref:ecomp,ref:rsecomp}, determines the degree to which the cathode emittance, given by:
\begin{eqnarray}
\epsilon_{n,x}=\frac{1}{mc}
\sigma_{x,0}\sigma_{p_x,0} = \sigma_x\sqrt{\frac{\mathrm{MTE}}{mc^2}},
\label{eqn:thermalemittance}
\end{eqnarray}
dominates the beam quality downstream.  Here $\sigma_{p_x,0}$ is the momentum variance intrinsic to the cathode material, which can be expressed in terms of the mean (kinetic) energry of the photoemitted electrons (MTE), and $\sigma_{x,0}$ is the spatial variance of the laser distribution.  
The rms emittance motivates a simple definition for the average transverse (normalized) brightness, defined generally as the particle flux per unit 4D transverse phase space volume \cite{ref:syncradps,ref:maxbb}:
\begin{eqnarray}
\overline{\mathcal{B}}_{n} = \frac{\bar{I}}{\epsilon_{n,x}\epsilon_{n,y}},
\label{eqn:avgbrightness}
\end{eqnarray}
where $\bar{I}$ is the average beam current.  

To characterize the contribution of the central core of the phase space to the emittance, as well as provide a pratical means to compare non-Gaussian beams, we define the emittance vs. fraction curve (see \cite{ref:syncradps,ref:lowemitter} for details): for every area in phase space $a$, we find a bounding contour $D(a)$ which maximizes the enclosed fraction $f$ of beam particles.   The rms emittance computed for the particles inside $D(a)$ defines the corresponding fractional emittance $\epsilon_{n,x}(f)$.  From this we define the core emittance as 
\begin{eqnarray}
\epsilon_{n,x}^{\mathrm{core}}=\left.d\epsilon/df\right|_{f=0} = \frac{1}{4\pi\rho_0},
\label{eqn:coredef}
\end{eqnarray}
where $\rho_0$ is the peak in the phase space distribution function (typically the centroid).  
The corresponding brightness as a function of fraction and peak brightness follow directly from Eqns.~(\ref{eqn:avgbrightness}-\ref{eqn:coredef}): 
\begin{eqnarray}
\overline{\mathcal{B}}_{n}(f) = \frac{\bar{I}f^2}{\epsilon_{n,x}(f)\epsilon_{n,y}(f)},\hspace{0.5cm}
\overline{\mathcal{B}}^{\mathrm{peak}}_{n}=\left.\overline{\mathcal{B}}_{n}\right|_{\mathrm{core}}.
\label{eqn:brightcore}
\end{eqnarray}
In addition to defining the principle figures of merit for high-brightness accelerators, Eqns.~(\ref{eqn:thermalemittance}-\ref{eqn:brightcore}) make clear the importance of preserving the thermal and cathode core emittances: the degree of conservation sets the extent to which the intrinsic cathode MTE determines the downstream beam quality.  For cathode emittance dominated beams, any improvement in the MTE translates into immediate improvement in the final beam quality.  With this property in mind, we turn to the main purpose of this work: demonstrating that the Cornell ERL injector, a 5-15 MeV machine featuring a DC gun followed by a short SRF linac, can produce beams with a high degree of emittance preservation in the beam dynamics regime set by next generation light sources.  

Originally designed to create low emittance, moderate bunch charge ($\leq$77 pC) beams at  high (1.3 GHz) repetition rate for a full hard x-ray ERL, the Cornell injector currently holds the world record for high average current from a photoinjector with cathode lifetimes suitable for an operating user facility \cite{ref:hcrecord}, as well as the record for lowest demonstrated emittance from a DC gun-based photoinjector at bunch charges of 19 and 77 pC \cite{ref:lowemitter}.  As of this work, the Cornell injector remains largely the same as described in \cite{ref:lowemitter}, with the most notable difference being the current operation of the DC gun at 395 kV.

For the measurements in this work, we used the LCLS-II injector 95\% normalized emittance and peak current specifications shown in Table-\ref{tab:lclsII_injector_params} as our working parameters \cite{ref:slacspecs1}.  
\begin{table}[htb]
\caption{LCLS-II Injector Specifications}\label{tab:lclsII_injector_params}
\begin{ruledtabular}
\begin{tabular}{ c | ccc}
Bunch Charge & 20 pC & 100 pC & 300 pC \\
\hline
95\% $\epsilon_{n,x,y}$ & 0.25 $\mu$m & 0.40  $\mu$m & 0.60 $\mu$m \\
Peak Current & 5 A & 10 A & 30 A
\end{tabular}
\end{ruledtabular}
\end{table}
For all direct phase space and longitudinal profile data taken with our Emittance Measurement System (EMS), we exclusively used a 50 MHz laser to limit the beam power deposited into the interceptive EMS diagnostics.
This laser system produces 520 nm, 1 ps rms pulses with comparable pulse energy  to the primary 1.3 GHz laser used for full repetition rate experiments \cite{ref:GHzlaser}.  Four rotatable birefringent crystals temporally shape the primary pulses by splitting each into 16 copies with tunable relative intensities set by the crystal rotation angles.  For transverse shaping, we used a beam expander and pinhole to clip the Gaussian laser distribution at roughly the half maximum intensity (truncation fraction of 50\%).  
All measurements were performed using a single NaKSb cathode with a $140\pm10$ meV MTE. 


%


In order to determine the injector settings that produce optimal emittances and peak currents, we ran Multi-Objective Genetic Algorithm (MOGA) optimizations using the 3D space charge code General Particle Tracer (GPT) \cite{ref:gpt1}.  For each of the LCLS-II nominal charges, the optimizer simultaneously minimized both the emittance and rms bunch length at the location of the EMS in the simulated injector, subject to realistic constraints on all relevant injector and beam parameters. 
For a detailed description of our 3D injector model, refer to \cite{ref:lowemitter}.  Additionally, we provided the optimizer with a realistic simulation of the laser distribution, and allowed the optimizer to vary the transverse pinhole size and truncation fraction, as well as the rotation of the longitudinal shaping crystal angles.  

The resulting Pareto fronts (shown later) provided injector settings that simultaneously satisfied both the 95\% emittance and peak current goals specified by the LCLS-II injector design.  In all cases, the optimizer chose a 9-9.5 MeV final beam energy at the EMS.  Additionally, the optimizer chose 0.73 mm, 1.9 mm, and 3.5 mm pinhole diameters, and roughly $50\%$ for the truncation fraction for the three bunch charges respectively.  The corresponding pinholes available at the time of measurement were 1 mm, 2 mm, and 3.5 mm.  Post processing of the optimized simulations showed a weak dependence of the transverse emittances on the temporal shaping crystal angles.  For simplicity, we tuned the crystal angles to produce a flat top temporally.  

For each bunch charge we loaded the corresponding optimal settings into the injector and tuned the machine to produce the lowest emittances possible while still meeting the peak current targets.  All critical machine parameters matched those chosen by the optimizer to within 5\%, with the notable exception of the pinhole used for the 20 pC measurements.  At these optimal machine settings, we measured the initial transverse laser distribution at the cathode, as well as the longitudinal electron current distribution, and both the horizontal and vertical projected phases spaces at the EMS.  From the phase spaces we computed the emittance vs. fraction curves, and the core emittances. 
The thermal emittances were computed directly from the measured transverse laser profiles according to Eqn.~(\ref{eqn:thermalemittance}).  In order to characterize the initial temporal laser shape, we measured the longitudinal electron beam current profile at near zero charge (0.02 $\pm$ 0.01 pC) with all RF cavities off.  Finally, we loaded the corresponding machine settings and measured laser distributions for each bunch charge into our virtual accelerator GUI \cite{ref:lowemitter}, and ran 250k macro-particle GPT simulations for comparison with measurement.

\begin{figure}[htb]
    \centering
    \includegraphics[width=80mm]{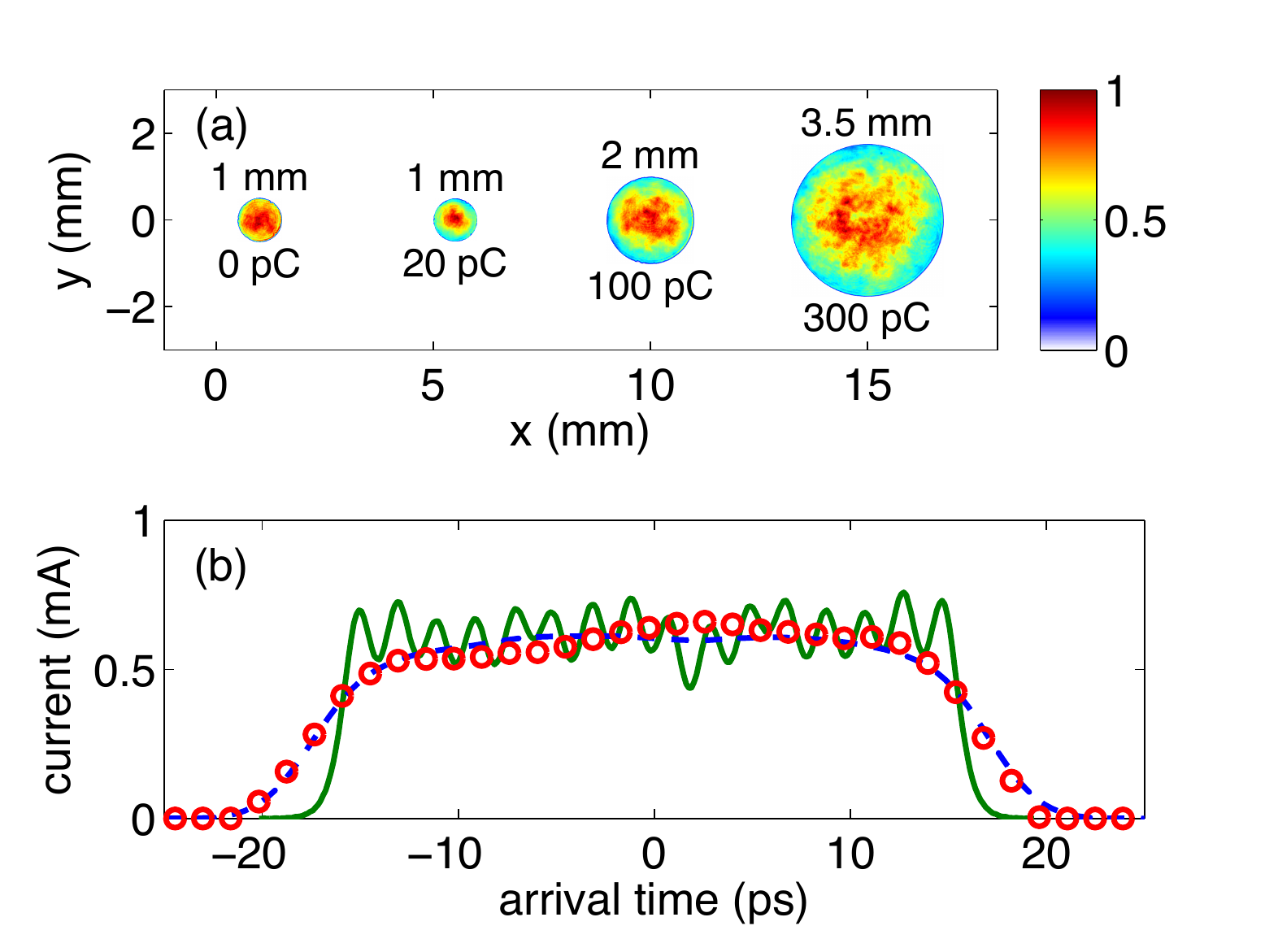}
    \caption{(a) The measured transverse laser distributions. (b) The simulated temporal laser distribution (green), the resulting electron current profile at the EMS from GPT (dashed blue), and the measured electron current profile (red).}
    \label{fig:laser_data}
\end{figure}

 Fig.~\ref{fig:laser_data} shows the measured laser distributions on a CCD camera located at the same distance from the clipping pinhole as the cathode.
To match the optimization results as best as possible, we tuned the laser spot size on the laser CCD so that the edge truncation fraction was $50\%$ using a beam expander.  Fig.~\ref{fig:laser_data}(b) shows the measured temporal current profile of the electron beam at the EMS (red), for a bunch charge of 0.02 $\pm$ 0.01 pC, and with all RF cavities off.  The green curve shows the simulated initial temporal laser distribution (normalized to 0.02 pC) and the resulting simulated electron beam current profile at the location of the EMS in GPT (dashed blue).

\begin{figure}[htb]
    \centering
    \includegraphics[width=80mm]{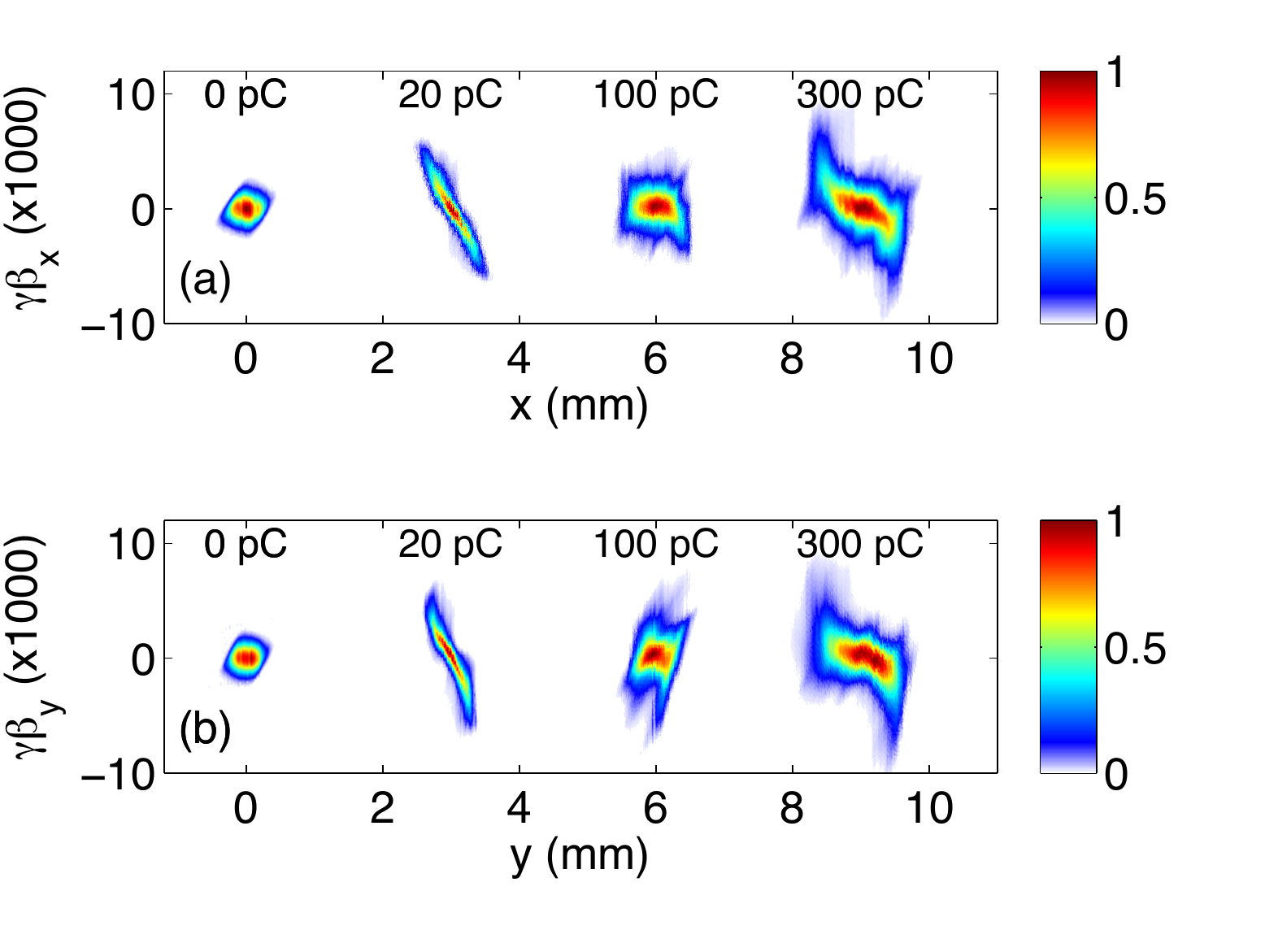}
    \caption{Measured horizontal (a) and vertical (b) projected phase spaces.}
    \label{fig:xy_ps_vs_q}
\end{figure}

Fig.~\ref{fig:xy_ps_vs_q} displays the measured horizontal and vertical projected phases spaces corresponding to the best measured emittances.  Note the use of the normalized mechanical momenta $\gamma\beta_{x_i} = p_{x_i}/mc$.  
One striking feature seen in Fig.~\ref{fig:xy_ps_vs_q} is the overall symmetry between the horizontal and vertical phase spaces.
\begin{figure}[tb]
    \centering
    \includegraphics[width=68mm]{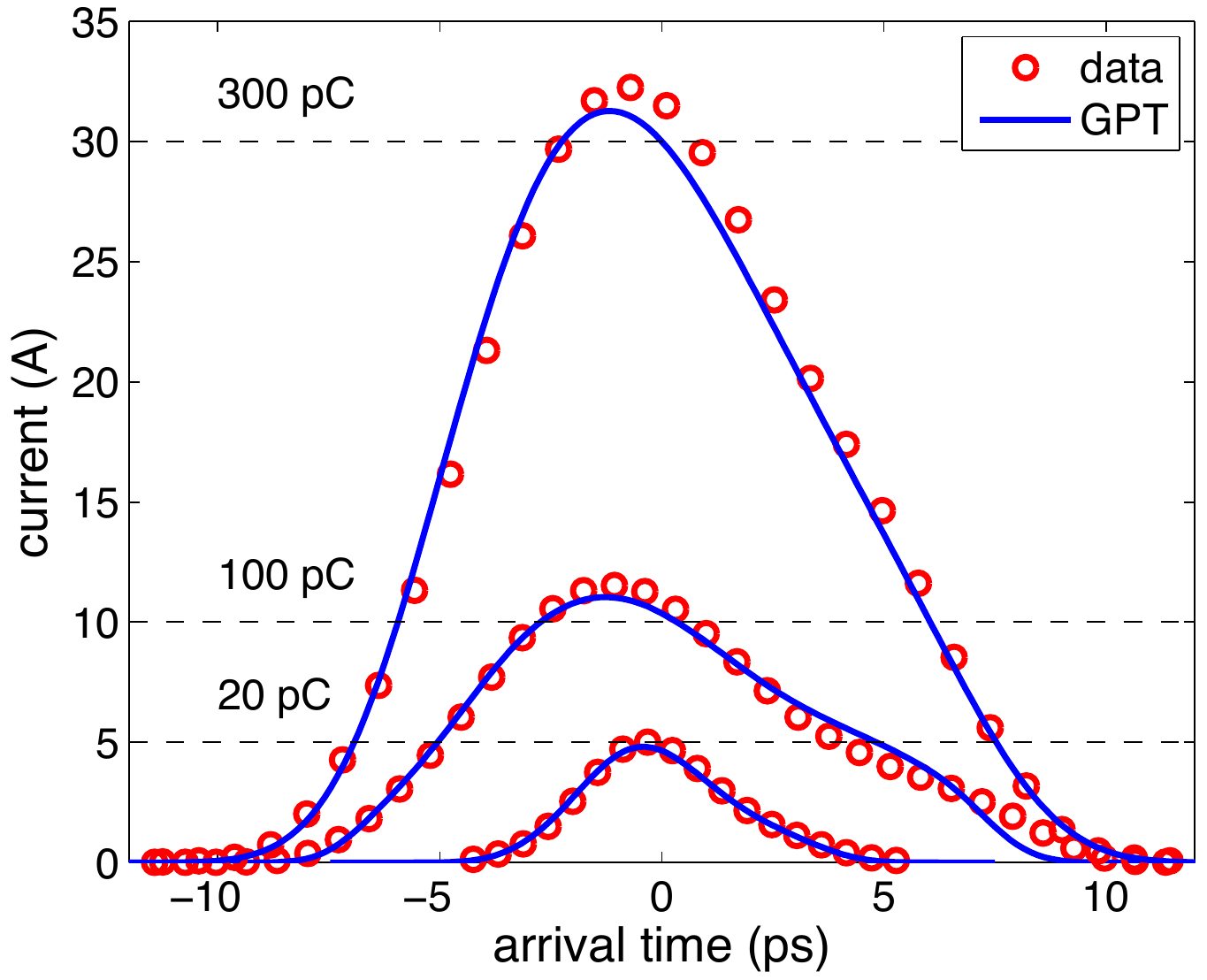}
    \caption{Comparison of the simulated (blue) and measured (red) current profiles as a function of bunch charge. Peak current targets are shown in black.}
    \label{fig:simcomp:current_profiles}
\end{figure}
Fig.~\ref{fig:simcomp:current_profiles} shows the comparison of the measured (red) and simulated longitudinal current profiles (blue).  In addition to the excellent agreement seen between measurement and simulation, we note that all peak current targets were met.

Table~\ref{tab:emittances} displays the thermal and core emittance at the cathode and the resulting measured 95\% (Table-\ref{tab:emitt}) and core emittances (Table-\ref{tab:coreemitt}) at the EMS.  
We estimate a $\pm6$\% relative error for the thermal emittances, and a $\pm$10\% relative error in the 95\% emittances measured with the EMS (up to the specified resolution of  $\leq$ 0.05 $\mu$m).
The random error between subsequent measurements using the EMS was typically $\leq$1\%.
\begin{table}
\caption{\label{tab:emittances} (a) Measured horizontal (vertical) thermal 95\% emittances at the EMS location.  (b) Initial and final measured horizontal (vertical) core emittances.}
\subtable[\hspace{0.2cm}Horizontal (vertical) projected emittance data.]{
\label{tab:emitt}
\begin{ruledtabular}
\begin{tabular}{c | ccc}
Charge & Thermal $\epsilon_{n}$  ($\mathrm{\mu m}$) & 95\% $\epsilon_{n}$ ($\mathrm{\mu m}$) & Ratio ($\%$)  \\ 
\hline
20 pC   & 0.12 (0.11) & 0.18 (0.19) & 67 (58) \\
100 pC & 0.24 (0.23) & 0.30 (0.32) & 80 (72) \\
300 pC & 0.42 (0.41) & 0.62 (0.60) & 67 (68)
\end{tabular}
\end{ruledtabular}
}\\
\subtable[\hspace{0.2cm}Horizontal (vertical) projected core emittance data.]{
\label{tab:coreemitt}
\begin{ruledtabular}
\begin{tabular}{c | ccc}
Charge & Cathode $\epsilon_{n,\mathrm{core}}$  ($\mathrm{\mu m}$) & EMS $\epsilon_{n,\mathrm{core}}$  ($\mathrm{\mu m}$) & Ratio ($\%$)  \\ 
\hline
20 pC    &  0.06 (0.06)  & 0.09 (0.08)  & 67 (75)  \\  
100 pC  &  0.14 (0.13)  & 0.16 (0.16)  & 85 (79)  \\
300 pC  &  0.26 (0.24)  & 0.30 (0.28)  & 87 (87)  
\end{tabular}
\end{ruledtabular}
}
\end{table}
We note that this data quantitatively reflects the qualitative symmetry seen in the phase space measurements (Fig.~\ref{fig:xy_ps_vs_q}), and as well as satisfies all of the LCLS-II injector emittance targets.  The table also shows the ratio of the thermal emittance and the final 95\% emittance, and the ratio of the initial and final core emittances.  In all measurements, the thermal emittances were preserved to within 58\%-80\%.  Similarly, the core emittances were preserved within 67-87\%.  We point out that the roughly 80-90\% preservation of the core emittance for all charges except 20 pC. In this case, the finite resolution of the EMS ($\leq$ 0.05 $\mu$m) likely becomes a contributing factor when measuring such small emittances. We conclude that the actual core emittance for this bunch charge is smaller than the quoted value, as suggested by simulation.  Nevertheless, these results demonstrate the main focus of this work: contrary to previous thought, DC gun based photoinjectors are capable of delivering cathode emittance dominated beams at high bunch charges suitable for use in next generation FELs like the LCLS-II.

In order to determine the effect of the laser shape on the emittances, we ran a second round of optimizations using the measured transverse laser distributions in Fig.~\ref{fig:laser_data}(a) and the crystal angles used to create the flattop Fig.~\ref{fig:laser_data}(b).  All other relevant injector parameters varied as before.  
\begin{figure}[tb]
    \centering
    \includegraphics[width=70mm]{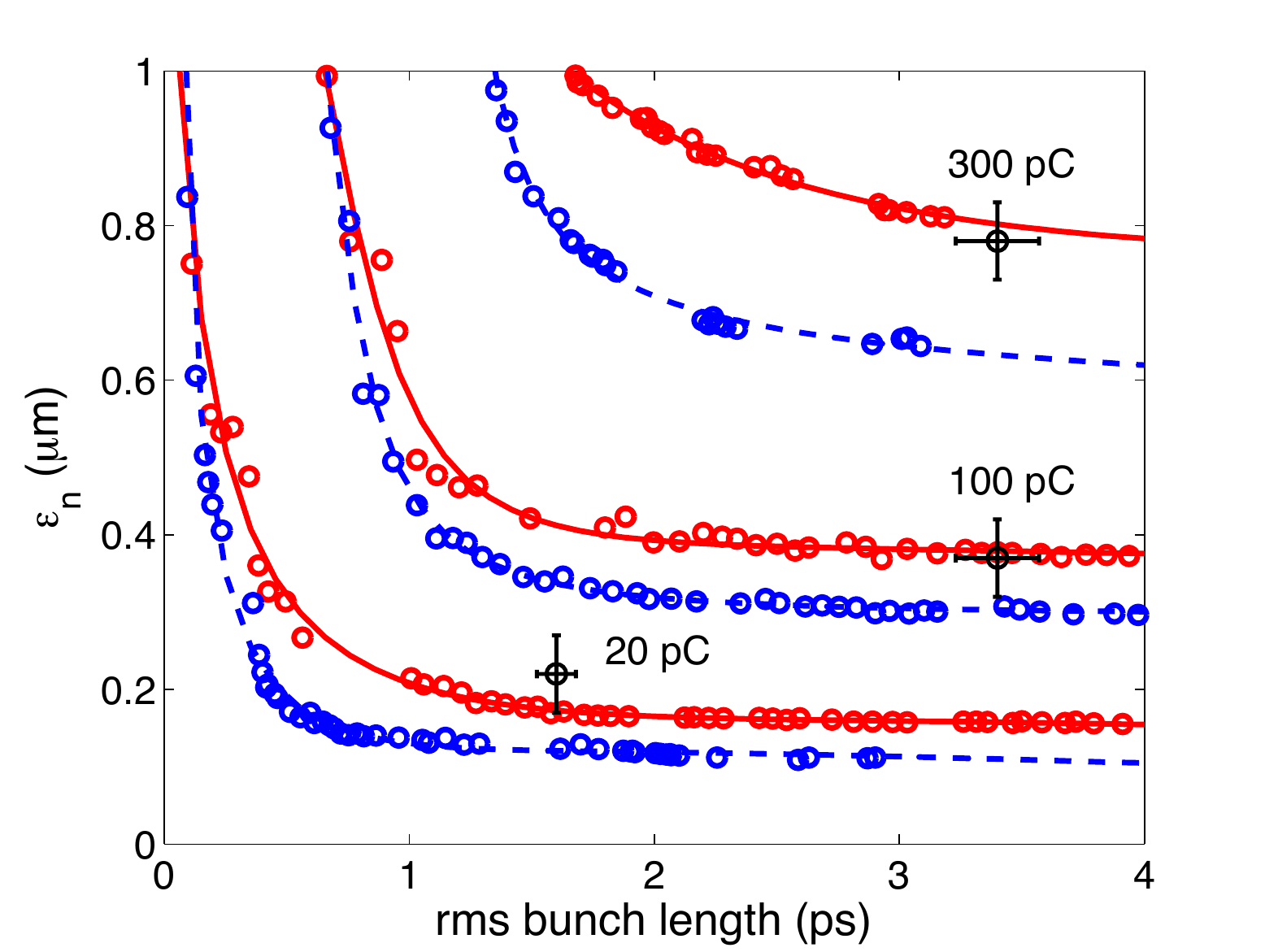}
    \caption{Optimized emittance vs. rms bunch length using (blue) a perfect variable truncated Gaussian and variable temporal distribution, (red) the measured laser distributions. Measured data are shown in black.}
    \label{fig:simcomp:enopt}
\end{figure}
Fig.~\ref{fig:simcomp:enopt} shows the \emph{average 100\% emittance}, $\epsilon_{n}=\frac{1}{2}(\epsilon_{n,x}+\epsilon_{n,y})$, vs. rms bunch length at each bunch charge.  Shown in blue are the initial optimizations with varied laser distribution parameters, and ideal transverse shape.  The red curves show the results of the second round of optimizations using the measured laser distributions (Fig.~\ref{fig:laser_data}).  The emittances corresponding to the data in Figs.~\ref{fig:xy_ps_vs_q}-\ref{fig:simcomp:current_profiles} and Table-\ref{tab:emittances} are shown in black.  We note that the emittance growth due to the laser (distance between blue and red curves at the measured bunch lengths shown in black) increases with bunch charge, as one might expect. 
For the 100 and 300 pC measurements, this produces roughly a 23\%, and 27\% relative emittance growth, due primarily to the error in transverse laser shape (as opposed to the pinhole size).  In the 20 pC case, the 42\% relative emittance growth seen is likely due to the use of a pinhole size 40\% larger than the optimal value.  

In this work, we have shown that optimal injector settings found using MOGA optimizations of 3D space charge simulations of the Cornell ERL injector produce machine states that preserve both the measured 95\% and core emittance, computed from direct phase space measurements, to within 57-87\% for 20, 100, 300 pC bunches.  Furthermore, the resulting measured emittances and longitudinal current profile show excellent agreement with corresponding GPT simulations, and meet the stated 95\% emittance and peak current specifications of the LCLS-II injector design.  Additionally, we have shown that the transverse laser shape plays an important role in determining the optimal emittances, adding further relevance to the recent demonstration of accurate, arbitrary transverse laser shaping at Cornell \cite{ref:lasershaping1,ref:lasershaping2}.  In conclusion, this work shows that DC gun based photoinjectors can produce cathode emittance dominated beams with single bunch beam quality rivaling that produced by RF gun based injectors for charges up to 300 pC, and represents a significant expansion of the beam dynamics regime for which DC gun-based injectors are applicable.

\begin{acknowledgments}
We acknowledge Jared Maxson for his useful discussions and interest in this work.
This work was supported, in part, by the LCLS-II Project and the US Department of Energy, Contract DE-AC02-76SF00515.
\end{acknowledgments}

\end{document}